\documentclass[aps,prd,twocolumn,showpacs,superscriptaddress,groupedaddress,ref,floatfix]{revtex4}  
\usepackage{graphicx}  
\usepackage{dcolumn}   
\usepackage{bm}        
\usepackage{amssymb}   
\usepackage{amsmath} 
\usepackage{natbib}
\usepackage{aas_macros} 
\usepackage{here} 

\begin{document}

\widetext


\title{Fast collective neutrino oscillations inside the neutrino sphere in core-collapse supernovae}

\author{Milad Delfan Azari}
\affiliation{Department of Pure and Applied Physics, Graduate School of Advanced Science and Engineering, Waseda University, 3-4-1 Okubo, Shinjuku, Tokyo 169-8555, Japan}

\author{Shoichi Yamada}
\affiliation{Department of Pure and Applied Physics, Graduate School of Advanced Science and Engineering, Waseda University, 3-4-1 Okubo, Shinjuku, Tokyo 169-8555, Japan}
\affiliation{Advanced Research Institute for Science and Engineering, Waseda University, 3-4-1 Okubo, Shinjuku, Tokyo 169-8555, Japan}

\author{Taiki Morinaga}
\affiliation{Department of Pure and Applied Physics, Graduate School of Advanced Science and Engineering, Waseda University, 3-4-1 Okubo, Shinjuku, Tokyo 169-8555, Japan}

\author {Hiroki Nagakura} 
\affiliation{Department of Astrophysical Sciences, Princeton University, Princeton, NJ 08544, USA}

\author {Shun Furusawa}
\affiliation{Department of Physics, Tokyo University of Science, Kagurazaka 1-3, Shinjuku, Tokyo 162-8601, Japan}

\author {Akira Harada}
\affiliation{Institute for Cosmic Ray Research, University of Tokyo, 5-1-5 Kashiwanoha, Kashiwa, Chiba 277-8582, Japan}

\author {\mbox{Hirotada Okawa}} 
\affiliation{Advanced Research Institute for Science and Engineering, Waseda University, 3-4-1 Okubo, Shinjuku, Tokyo 169-8555, Japan}
\affiliation{Yukawa Institute for Theoretical Physics, Kyoto University, Oiwake-cho, Kitashirakawa, Sakyo-Ku, Kyoto, 606-8502, Japan}
\affiliation{Waseda Institute for Advanced Study, 1-6-1 Nishi Waseda, Shinjuku, Tokyo 169-8050, Japan}

\author {\mbox{Wakana Iwakami}} 
\affiliation{Advanced Research Institute for Science and Engineering, Waseda University, 3-4-1 Okubo, Shinjuku, Tokyo 169-8555, Japan}
\affiliation{Yukawa Institute for Theoretical Physics, Kyoto University, Oiwake-cho, Kitashirakawa, Sakyo-Ku, Kyoto, 606-8502, Japan}

\author {Kohsuke Sumiyoshi} 
\affiliation{National Institute of Technology, Numazu College, Ooka 3600, Numazu, Shizuoka 410-8501, Japan}

\date{\today}

\begin{abstract}

Neutrinos are believed to have a key role in the explosion mechanism of core-collapse \mbox{supernovae} as they carry most of the energy released by the gravitational collapse of a massive star. If their flavor is converted fast inside the neutrino sphere, the supernova explosion may be influenced. This paper is reporting the results of the extended work of our previous paper. We perform a thorough survey of the ELN crossing in one of our self-consistent, realistic \mbox{Boltzmann} simulations in two spatial dimensions under axisymmetry for the existence of the \mbox{crossings} between $\nu_e$ and $\bar\nu_e$ angular distributions, or the electron lepton number (ELN) crossing. We report for the first time the positive detections deep inside the core of the massive star in the vicinity of neutrino sphere at \mbox{$r$ $\approx$ 16 - 21 km.} We find that low values of the electron fraction $Y_e$ produced by convective motions together with the appearance of light elements are critically important to give rise to the ELN crossing by enhancing the chemical potential difference between proton and neutron, and hence by mitigating the Fermi-degeneracy of $\nu_e$. Since the region of positive detection are sustained and, in fact, expanding with time, it may have an impact on the explosion of core-collapse supernovae, observational neutrino astronomy and nucleosynthesis of heavy nuclei. 
     
\end{abstract}

\maketitle{}

\section{introduction}

Neutrinos ($\nu^,$s) are fermions and are one of the most abundant particles in the universe \cite{Kostelecky:1995xc}. They are massive particles, with their mass eigenstates being not diagonal with their flavor eigenstates \cite{1998PhRvL..81.1562F} and, as a result, they oscillate among their three flavors ($\nu_e$, $\nu_{\mu}$ and $\nu_{\tau}$) while propagating in vacuum \cite{2008PhRvD..78h3007G}. When neutrinos propagate through a medium, they gain effective mass due to interactions with matter and give rise to a phenomena called the \mbox{Mikheyev-Smirnov-Wolfenstein} (MSW) effect \cite{Mikheev:1987qk,1978PhRvD..17.2369W,2008PhRvD..78h3007G}. The self-energy can be also generated by interacting with other neutrinos and if the environment is neutrino-rich, the so-called collective neutrino oscillations occur in which the flavor evolution becomes nonlinear and hence very complicated \cite{2010ARNPS..60..569D,2008PhRvD..78h5012E,1992PhLB..287..128P,2007PhRvD..75h3002R}.

Core-collapse supernovae (CCSNe), which are the end phase in the evolution of the massive stars with a zero-age-main-sequence (ZAMS) mass of $\gtrsim$ 8 $M_\odot$, are one of the most energetic explosions in the universe and are environments where an enormous amount of neutrinos are produced \cite{Mirizzi:2015eza}. Although the exact explosion mechanism of CCSNe is not fully understood, it is well known that neutrinos are the key players as almost all of the energy released by the gravitational collapse of a massive star is emitted in the form of neutrinos and the kinetic energy of ejected materials in the explosion is only about one percent of the neutrino energy \cite{Janka:2017vlw}.
      
About a decade ago, it was pointed out by Sawyer that the neutrino flavor conversion may occur near the neutrino sphere and if true, it will have a strong impact on the explosion mechanism of CCSNe. He reported in a series of papers \cite{2005PhRvD..72d5003S,2009PhRvD..79j5003S,2016PhRvL.116h1101S} a mechanism called "fast oscillation", in which the frequency is proportional to the neutrino potential $\mu \sim \sqrt2 G_{F}n_{\nu}$ . It is known that the fast flavor conversions occur when the "electron-lepton-number (ELN) crossing" exists \cite{2016PhRvL.116h1101S,Dasgupta:2017oko,Capozzi:2017gqd,Dasgupta:2016dbv,Chakraborty:2016lct,2017PhRvL.118b1101I}, that is, the difference between the energy-integrated distribution functions of electron-type neutrinos and their anti-particles changes its sign as a function of propagation direction. It is believed that CCSNe are one of the best astrophysical environments for the fast flavor conversion because neutrinos with different flavors are highly populated there, having different angular distributions. 

So far the fast flavor conversions in the realistic settings have been studied only in 1D under spherical symmetry and no sign of the ELN crossing has been found \cite{Tamborra:2017ubu}. In our previous paper \cite{Azari:2019jvr}, we conducted a pilot study based on a small number of data extracted from our \mbox{fully self-consistent} realistic simulations of CCSNe in two spatial and three momentum dimensions with our Boltzmann-neutrino-radiation-hydrodynamics code \cite{Nagakura:2017mnp}. We did not find any crossing between $\nu_e$ and $\bar\nu_e$ angular distributions at a specific point (\mbox{$r$ = 44.8 km,} \mbox{$\theta$ = 2.36 rad)} in three different time-steps after bounce (15, 190 and 275 ms). In contrast with our results, a positive detection of the ELN crossings at $r$ $\gtrapprox$ 50 - 70 km was reported in \cite{Abbar:2018shq}. It should be mentioned that their results were based on the time-independent neutrino distributions computed for some fixed matter profiles and hence were not fully self-consistent. Very recently Nagakura et al. \cite{Nagakura:2019sig} also found the crossings at similar regions in one of their latest self-consistent simulations with an updated EOS.

The goal of this paper is to conduct a more thorough survey of the ELN crossing in our fully self-consistent, realistic simulations of CCSN in two dimensions under axisymmetry with our \mbox{Boltzmann-neutrino-radiation-hydrodynamics} code that computes neutrino transport together with hydrodynamics. The same data as employed in the previous study [21] are again used. We report for the first time in the following a positive result found at about \mbox{$r \gtrapprox$ 16 - 21 km}, which is inside the neutrino sphere. If the fast flavor conversion occurs at this small radius indeed, it is likely to have some impact on the supernova explosion.

This paper is structured as follows. In Sec.II, we summarize the equations to give the dispersion relation for the fast collective neutrino oscillations, which will be used in the later analysis, and present briefly the numerical models adopted in this study. Sec.III presents the survey results and finally in Sec.IV we conclude the paper with some discussions.   

\section{Method and models}
\subsection{Dispersion relation}

In this paper, we rely on the ELN crossing as a criterion of the fast flavor conversion and do not conduct linear analysis as in the previous paper \cite{Azari:2019jvr} in most cases. We will still use, however, in some cases the dispersion relation and the growth rate of the fast conversion for detailed analysis. We hence give the procedure to obtain the dispersion relation. Following the previous works \cite{Sigl:1992fn,Banerjee:2011fj,2017PhRvL.118b1101I,Azari:2019jvr,Morinaga:2018aug,Strack:2005ux,Volpe:2015rla,Hansen:2016klk}, neglecting ordinary collisions, we begin with the equation of motion for density matrix  $\rho$ as

 \begin{equation}
(\partial_{t}+\mathbf{v}\cdot\boldsymbol{\nabla}_{\boldsymbol{r}})\rho=i[\rho,H].\label{freestreaming}
\end{equation}
The Hamiltonian H in Eq. (\ref{freestreaming}) is written as 
\begin{equation}
H = H_{\mathrm{vacuum}}+ H_{\mathrm{matter}}+ H_{\mathrm{collective}},\label{Hamiltonian}
\end{equation}
where each term is given as 

\begin{eqnarray}
H_{\mathrm{vacuum}} &=& \dfrac{M^{2}}{2E},\\
H_{\mathrm{matter}} &=& v^{\mu}\Lambda_{\mu}\dfrac{\sigma_{3}}{2},\\
H_{\mathrm{collective}} &=& \sqrt{2}G_{F}\int \frac{E'^2dE'}{2\pi^2}d\Upsilon'v^{\mu}v_{\mu}'\rho'.
\end{eqnarray}
In the above expressions, $M^2$ = diag ($m^2_1,m^2_2,m^2_3$) in the mass eigenstates is the mass-squared matrix in vacuum. In $H_{\mathrm{matter}}$, we define the four velocity of neutrino as \mbox{$v^{\mu}=(1,\mathbf{v})$}; $\Lambda^{\mu}=\sqrt{2}G_{F}(n_{e}-n_{e^+})$ $u^{\mu}$ is the matter potential given with four velocity of matter $u^{\mu}$; $\sigma_3$ is the third Pauli matrix $\sigma_3$ = $\begin{pmatrix}1 & 0\\
0 & -1 
\end {pmatrix}$. In this paper the metric is assumed to be diag(1, -1, -1, -1).
 In $H_{\mathrm{collective}}$, which is responsible for the collective neutrino oscillation, $G_F$ is the Fermi coupling constant, $\rho'$ denotes the density matrix for the neutrinos with energy $E'$, which moves at the four velocity of \mbox{$v'^{\mu}$ = $(1, \mathbf{v'})$;} $d{\Upsilon'} = d\mathbf{v'}/4\pi$ is the infinitesimal solid angle in momentum space normalized by $4\pi$. 

As in our previous paper \cite{Azari:2019jvr}, we work in the two-flavor ($\nu_{e}$ and $\nu_{x}$) approximation as a common practice for simplicity, where $\nu_{x}$ stands for $\nu_{\mu}$ and $\nu_{\tau}$ collectively. 
The density matrix in Eq. (\ref{freestreaming}) is expressed as  
\begin {equation}
\rho=\dfrac{f_{\nu_{e}}+f_{\nu_{x}}}{2}+\dfrac{f_{\nu_{e}}-f_{\nu_{x}}}{2}\begin{pmatrix}s_{\mathbf p} & S_{\mathbf p}\\
S_{\mathbf p}^{*} & -s_{\mathbf p}\label{density}
\end{pmatrix},
\end{equation}
where the off-diagonal element in the matrix, $S_{\mathbf{p}}(t,\mathbf r)$, is the complex scalar field and represents the flavor coherence between $\nu_e$ and $\bar{\nu}_e$ for momentum $\mathbf p$; the diagonal element, $s_{\mathbf{p}}(t,\mathbf r)$, is the real field and obeys $\left|{S_{\mathbf p}}^2\right|$+$s_{\mathbf p}^2$ =1; $f_{\nu_{e}}$ and $f_{\nu_{x}}$ are the neutrino distribution functions for $\nu_{e}$ and $\nu_{x}$, respectively. When the neutrino is in one of the flavor eigenstates, $s_{\mathbf p}$ and $S_{\mathbf p}$ are $1$ and $0$, respectively, and Eq. (\ref{density}) can be written as
\begin {equation}
\rho=\dfrac{f_{\nu_{e}}+f_{\nu_{x}}}{2}+\dfrac{f_{\nu_{e}}-f_{\nu_{x}}}{2}\begin{pmatrix}1 & 0\\
0 & -1
\end{pmatrix}.
\end{equation}
We assume $\nu_x$ and $\bar\nu_x$ have the same distributions in this paper. As we focus on the fast modes here, we ignore $H_{\mathrm {vacuum}}$ in the following. Then the equations no longer include $E$ explicitly and we will deal with the \mbox{energy-integrated} quantities, e.g., $S_{\mathbf v}$ = $\int E^2 dE S_{\mathbf p}$ from this point on. By assuming that $S_{\mathbf v}\ll1$, we linearize \mbox{Eq. (\ref{freestreaming})} integrated over energy and obtain the following EOM \mbox{for $S_{\mathbf v}$}: 
\begin{equation}
i(\partial_{t}+\mathbf{v}\cdot\boldsymbol{\nabla}_{\mathbf{r}})S_{\mathbf{v}}=v^{\mu}(\Lambda_{\mu}+\Phi_{\mu})S_{\mathbf{v}}-\int d\Upsilon' v^{\mu}v_{\mu}'G_{\mathbf{v'}}S_{\mathbf{v'}}\label{EQNLIN},
\end{equation}
where $G_{\mathbf{v}}$ is the electron-lepton number (ELN) angular distribution defined as 
\begin{equation}
G_{\mathbf{v}}=\sqrt{2}G_{F}\int_{0}^{\infty}\frac{dEE^{2}}{2\pi^{2}}\left[f_{\nu_{e}}(E,\mathbf{v})-f_{\bar{\nu}_{e}}(E,\mathbf{v})\right].\label{gv}
\end{equation}
The corresponding ELN current is $\Phi^{\mu} \equiv\int\frac{d\mathbf{v}}{4\pi}G_{\mathbf{v}}v^{\mu}.$ If the solution of Eq.~(\ref{EQNLIN}) is assumed to take the form of \mbox{$S_{\mathbf{v}}=Q_{\mathbf{v}}e^{-i(\Omega t-\mathbf{K}\cdot\mathbf{r})}$}, the equation for the amplitude $Q_{\mathbf{v}}$ is given as 
\begin{equation}
v^{\mu}k_{\mu}Q_{\mathbf{v}}=-\int d\Upsilon^{'}v^{\mu}v_{\mu}'G_{\mathbf{v}'}Q_{\mathbf{v}'},\label{EigenEq2}
\end{equation}
where $k^{\mu}=K^{\mu}-\Lambda^{\mu}-\Phi^{\mu}$ with $k^{\mu}=(\omega,\mathbf{k})$ and \mbox{$K^{\mu}=(\Omega,\mathbf{K})$}.

Equation (\ref{EigenEq2}) can be recast into 
\begin{equation}
Q_{\mathbf{v}} = v^{\mu}a_{\mu}/v^{\mu}k_{\mu}\label{Qv}  
\end{equation}
with $a_{\mu}\equiv-\int\frac{d\mathbf{v}}{4\pi}v_{\mu}G_{\mathbf{v}}Q_{\mathbf{v}}$. Putting Eq. (\ref{Qv}) into \mbox{Eq. (\ref{EigenEq2})}, we obtain 
\begin{equation}
v_{\mu}\Pi^{\mu\nu}(\omega,\mathbf{k})a_{\nu}=0\label{PI},
\end{equation}
where the polarization tensor $\Pi^{\mu\nu}$ is given as 
\begin{eqnarray}
\Pi^{\mu\nu} & =&\eta^{\mu\nu}+\int\frac{d\mathbf{v}}{4\pi}G_{\mathbf{v}}\frac{v^{\mu}v^{\nu}}{\omega-\mathbf{v}\cdot\mathbf{k}}\nonumber \\
\end{eqnarray}
with the Minkowsky metric $\eta^{\mu\nu}$ = diag(1,-1,-1,-1). 
Equation (\ref{PI}) has non-trivial solutions if and only if 
\begin{equation}
\det\Pi=0.\label{DR}
\end{equation}

This equation gives us the dispersion relation between $\omega$ and $\mathbf{k}$. As shown in our previous study \cite{Azari:2019jvr}, it depends on the direction of the wave vector $\mathbf{k}$ in general and the radial direction is not necessarily the most important direction.

\begin{figure*}[t]
\begin{tabular}{ccc}
\includegraphics[width=4cm]{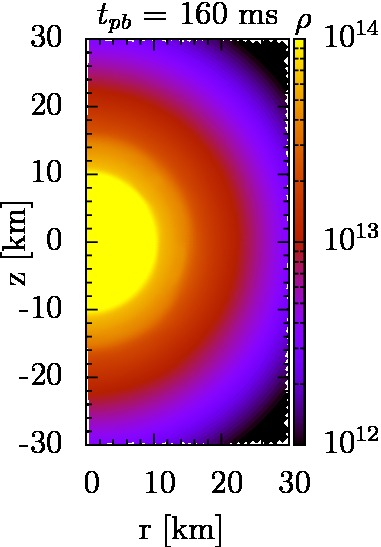}\includegraphics[width=4cm]{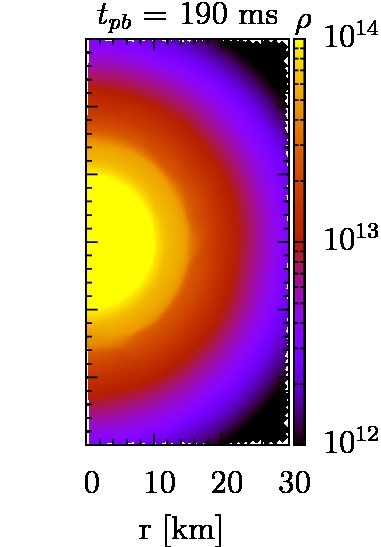}\includegraphics[width=4cm]{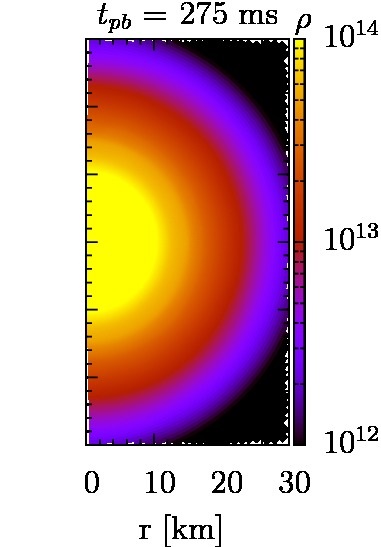}\\
\includegraphics[width=4cm]{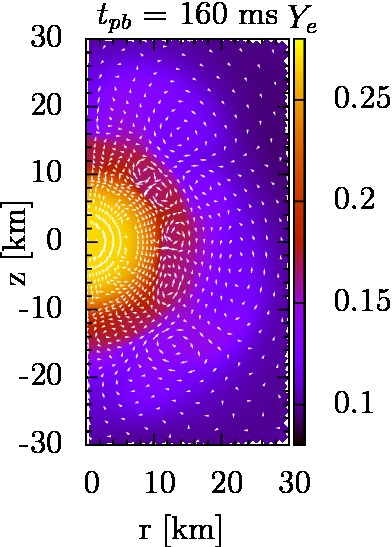}\includegraphics[width=4cm]{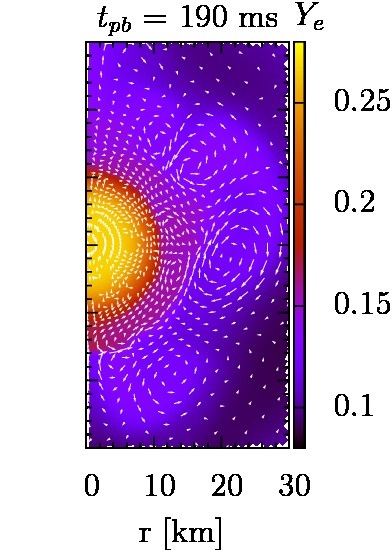}\includegraphics[width=4cm]{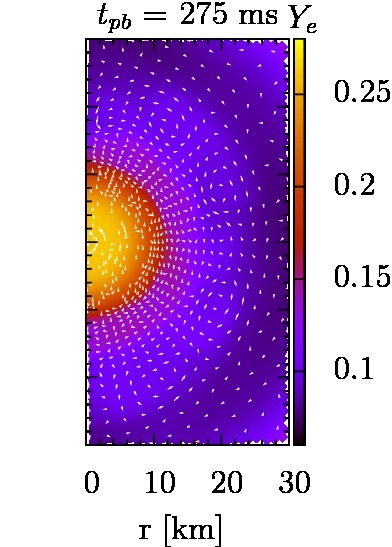}\\
\includegraphics[width=4cm]{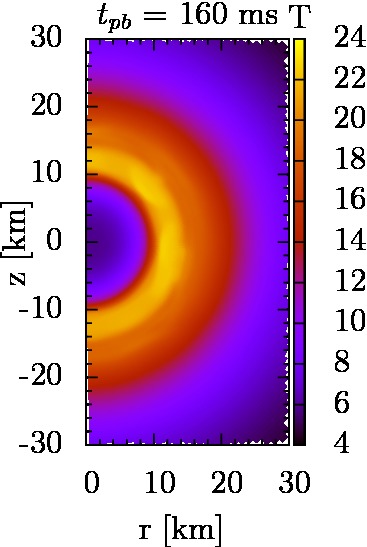}\includegraphics[width=4cm]{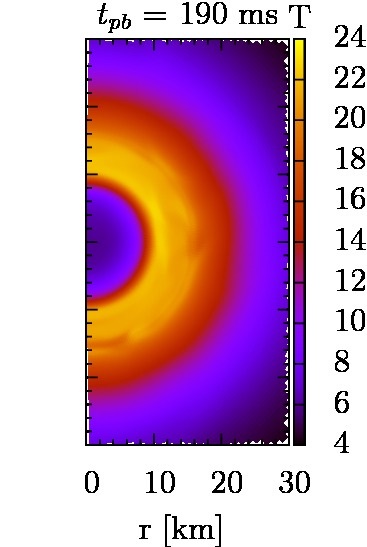}\includegraphics[width=4cm]{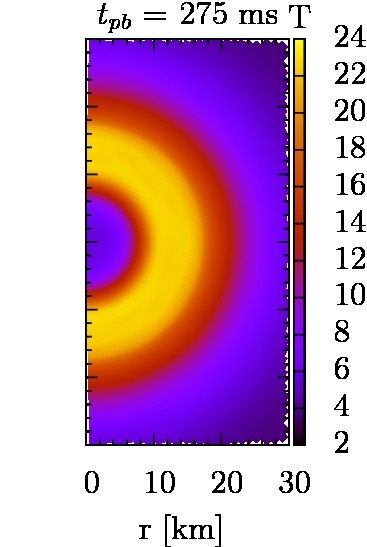}\\
\end{tabular} \caption{\label{density} The density (top), $Y_e$ (middle) and temperature (bottom) distributions in the half meridian section at three different post-bounce times: $t_{pb}$ = 160 ms (left), 190 ms (center) and 275 ms (right). In the middle row the arrows show the matter velocities.} 
\end{figure*}

\subsection{Models}

As in our previous work \cite{Azari:2019jvr}, we conduct our analysis to the results of the realistic two-dimensional (2D) fully self-consistent \mbox{Boltzmann-neutrino-radiation-hydrodynamics} simulations for the non-rotating progenitor model of \mbox{11.2 $M_{\odot}$} \cite{2002RvMP...74.1015W}, which were performed on the Japanese K supercomputers \cite{Nagakura:2017mnp}. In these simulations, three neutrino species, $\nu_e$, $\bar\nu_e$ and $\nu_x$ are considered and their distributions are computed on spherical coordinates ($r$, $\theta$) under spatial axisymmetry; we also employ spherical coordinates in momentum space \mbox{($\textit{E}$, $\theta_{\nu}$, $\phi_{\nu}$)}, in which the two angles are measured from the local radial direction. The computational domain covers $0 \leq r \leq 5000$ km, \mbox{0 $\leq\theta\leq \pi$}, \mbox{0 $\leq \textit{E}$ $\leq$ 300 MeV}, 0 $\leq$ $\theta_{\nu}$ $\leq$ $\pi$ and 0 $\leq$ $\phi_{\nu}$ $\leq$ 2$\pi$ with 384($r$), 128($\theta$), 20($\textit{E}$), 10($\theta_{\nu}$) and 6($\phi_{\nu}$) mesh cells. The results of the Furusawa-Shen equation of state (FSEOS), which is based on the relativistic mean field theory for nuclear matter \cite{Furusawa:2011wh,Furusawa:2013rta}, are adopted also in this paper, simply for continuity from the previous pilot study \cite{Azari:2019jvr}.

In this work, unlike in our previous paper \cite{Azari:2019jvr}, which we analyzed a single point (\mbox{$r = 44.8$km,} \mbox{$\theta$ = 2.36 rad)} alone, we make a survey for all the grid points inside the shock wave, searching for the crossings in the angular distributions between the electron-type neutrinos $\nu_e$ and anti-electron neutrinos $\bar\nu_e$, i.e., the change of sign in the ELN defined as
\begin{equation}
ELN (\theta_{\nu}, \phi_{\nu}) = \int\frac{E^2dE}{2\pi^2} (f_{\bar\nu_e} (E, \theta_{\nu}, \phi_{\nu}) - f_{\nu_e} (E, \theta_{\nu}, \phi_{\nu}))  ,\label{cross}
\end{equation}
where $f_{\bar\nu_e}$ and $f_{\nu_e}$ are the neutrino distribution functions of $\bar\nu_e$ and $\nu_e$, respectively. We also look into the dispersion relation at some representative points.
\begin{figure*}[t]
\begin{tabular}{cc}
\includegraphics[width=10cm]{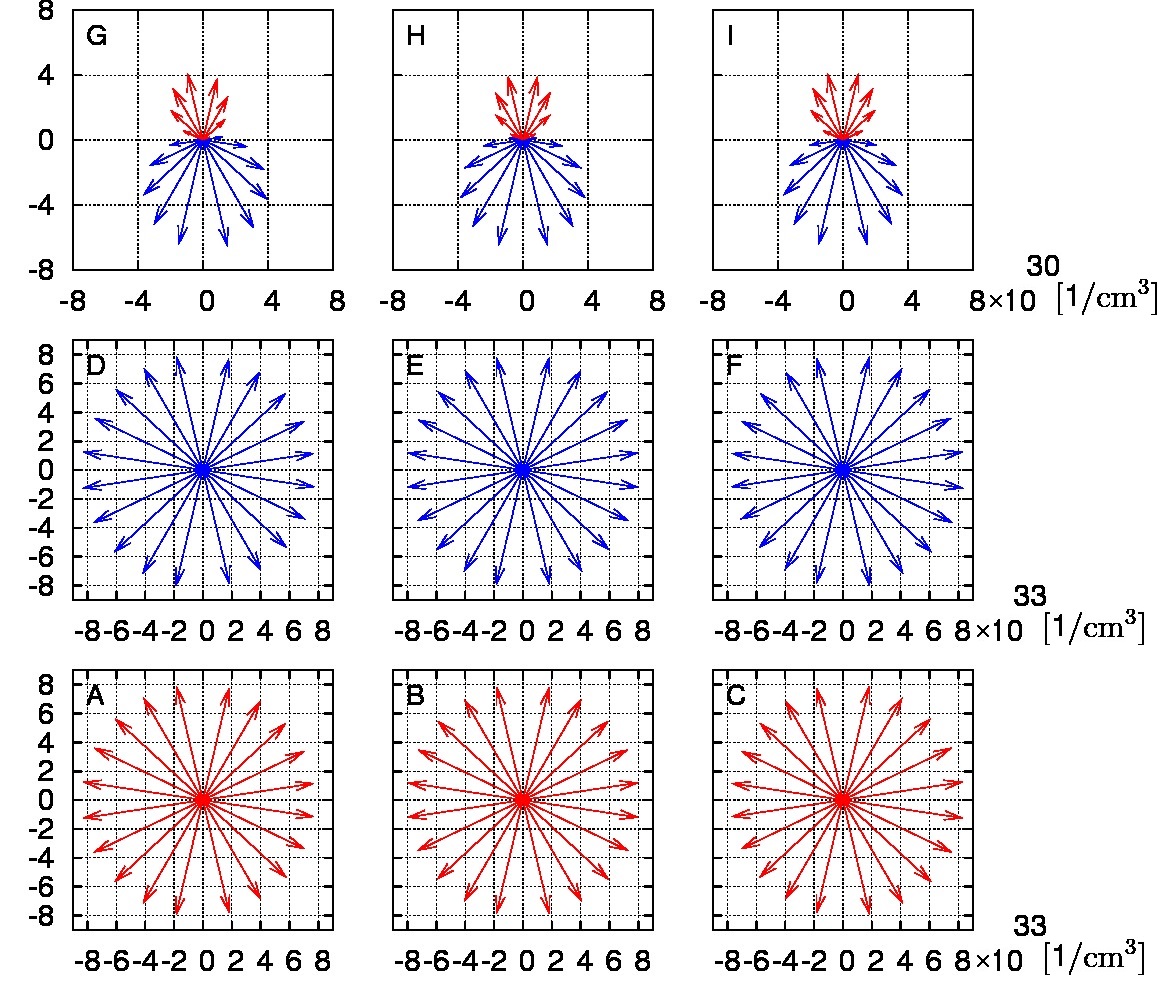}
\end{tabular} \caption{\label{ang190} Energy-integrated angular distributions of $\nu_e$ (bottom row) and $\bar\nu_e$ (middle row) and their difference (top row) on three meridian sections in momentum space at $t_{pb}$ = 190 ms. The red (blue) color represent $\nu_e$ ($\bar{\nu}_e$) for the bottom and middle rows whereas it implies that $\nu_e$ ($\bar{\nu}_e$) is dominant on the top row. The left, center and right columns display the meridian sections for the pairs of \mbox{$\phi_{\nu}$[radian] = (0.35, 3.49),} (1.57, 4.71) and (2.78, 5.92), respectively.} 
\end{figure*} 

\begin{figure*}[t]
\begin{tabular}{cccc}
\includegraphics[width=3.4cm]{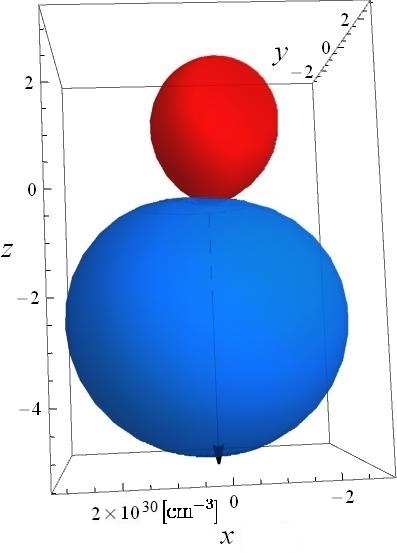}\includegraphics[width=4cm]{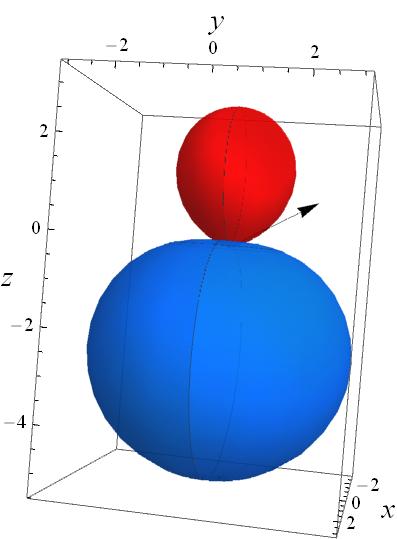}\includegraphics[width=4cm]{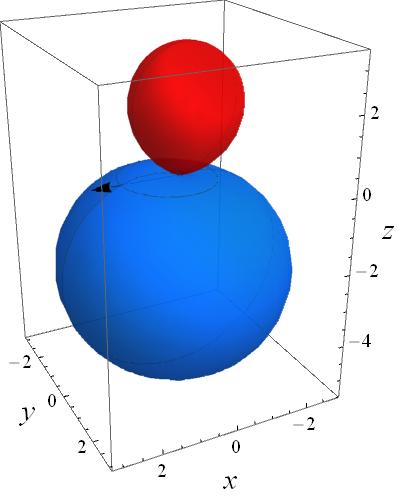}\includegraphics[width=4cm]{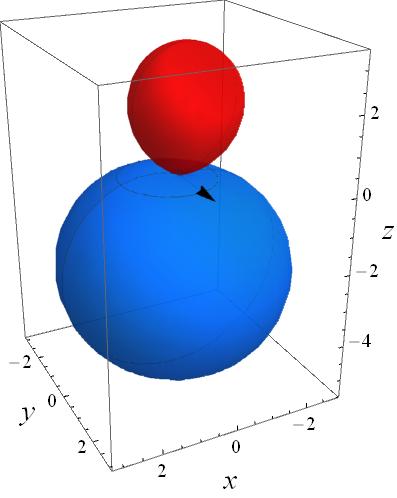}\\
\includegraphics[width=4cm]{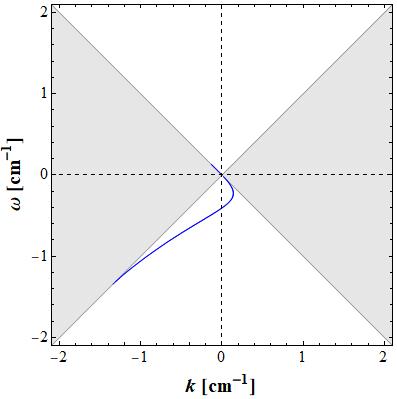}\includegraphics[width=4cm]{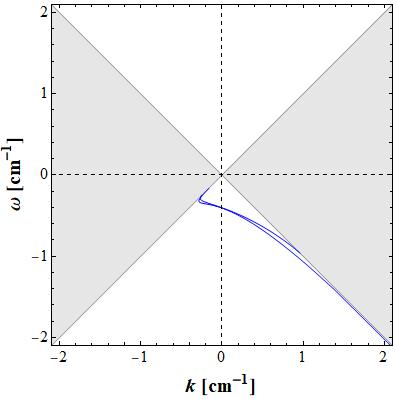}\includegraphics[width=4cm]{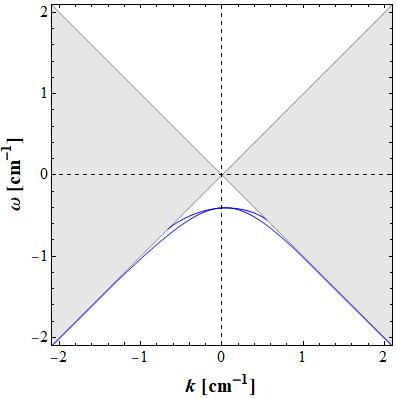}\includegraphics[width=4cm]{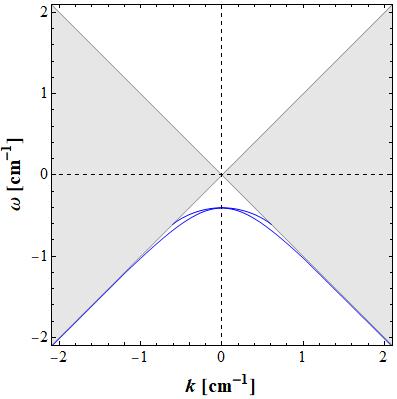}\\ 
\end{tabular} \caption{\label{nu190} Angular distribution differences between $\nu_e$ and $\bar\nu_e$ (upper panels) and the DRs (lower panels) for different wave vectors $\mathbf k$ of perturbations given in the upper panels at \mbox{$t_{pb}$ = 190 ms} for \mbox{$r$ = 16.5 km} and $\theta$ = 2.1 rad. In the upper panels, ELN defined in Eq. (14) is shown as a surface with the red (blue) color indicating the dominance of $\nu_e$ ($\bar{\nu}_e$). The z-axis corresponds to the local radial direction. Black arrows indicate the direction of wave vector \textbf{k}. In the bottom panels, the blues lines represent different branches of stable modes.} 
\end{figure*}

Figure \ref{density} is the contour plots for \mbox{density $\rho$}, electron fraction $Y_{e}$ and temperature $T$, respectively. White arrows represent the matter velocity, in the central portion ($r\lesssim$ 30 km) of the core, on which we will focus in this paper. These are the data from the same snapshots at $t_{pb}$= 190 and 275 ms as used in our previous paper. Although we have explored various post-bounce times this time, we will mainly adopt the results at these times for reasons that will become clear.

\section{results}

We report here the positive detection of the crossing in the region of $r \sim$ 16 - 18 km at \mbox{$t_{pb}$ = 190 ms}. These crossings are sustained for later times to \mbox{$t_{pb}$ = 275 ms}, at which time they occur in even wider region of \mbox{$r \sim$ 16 - 21 km}.  It is important that regions are located in the vicinity of the neutrino sphere and, as we mentioned already, the fast flavor conversion there may have an influence not only on the terrestrial observations of supernova neutrinos and nucleosyntheses but also on the supernova explosion itself.

In order to do further analysis, we pick up a point at the radius \mbox{$r$ = 16.5 km} and the spatial \mbox{zenith $\theta$ = 2.1 rad} in the crossing region for \mbox{$t_{pb}$ = 190 ms} as a representative. We show in Fig.~\ref{ang190} the energy-integrated angular distributions of $\nu_e$ (bottom row), $\bar\nu_e$ (middle row) and their difference (top row) in the meridian sections of momentum space for different values of the azimuthal angle $\phi_{\nu}$. The colors indicate neutrino species: red and blue arrows represent $\nu_e$ and $\bar\nu_e$, respectively for the bottom and middle rows; they indicate which species is dominant for the top row. The length of each arrow indicates the value of the distribution function for the bottom and middle rows and its difference between the two species for the top row in the direction that the arrow specifies. The vertical direction corresponds to the local radial direction. Each column shows the meridian section which corresponds to a pair of azimuthal angles in momentum space: panels A, B and C have \mbox{$\phi_{\nu}$[radian] = (0.35, 3.49),} (1.57, 4.71) and \mbox{(2.78, 5.92),} respectively. Note that the scales are almost the same in these panels, which is in sharp contrast with our previous results \cite{Azari:2019jvr}, where $\nu_e$ overwhelms $\bar{\nu}_e$ by a large factor. One can see that both distributions are almost isotropic. However, some differences, albeit small, do exist, which give rise to the crossing. Indeed, $\nu_e$ is still dominant in the radially outward direction, $\bar{\nu}_e$ prevails in the opposite direction as shown on the top row. It is also recognized that the distributions are not axisymmetric.

We think that the small differences in the angular distributions between $\nu_e$ and $\bar\nu_e$ are not numerical artifacts. Firstly, the neutrino distributions in momentum space are almost isotropic without fine features on small angular scales in the region of our concern as mentioned. This implies that many angular mesh points are not necessary. Secondly, we already studied the accuracy of our simulations with the same angular resolution in other papers (see, for example,\cite{Nagakura:2019rdf} and \cite{Richers:2017awc}) and found that it is very good in this region.  

The angular distribution differences between $\nu_e$ and $\bar\nu_e$ are also exhibited in the upper row of Fig.~\ref{nu190}. The surfaces in these 3D plots are the ELN defined in Eq. (14) as a function of the propagation direction. The blue part corresponds to the directions, in which $\bar\nu_e$ is dominant over $\nu_e$ and the opposite is true for the red section. The z-axis is aligned with the local radial direction and the x-axis is chosen so that it is contained in the spatial meridian section through that point. The four panels give the same angular distribution but viewed from different directions. It is clear that $\bar\nu_e$ overwhelms $\nu_e$ in the radially inward directions.

\begin{figure}[t]
\begin{tabular}{c}
\includegraphics[width=8cm]{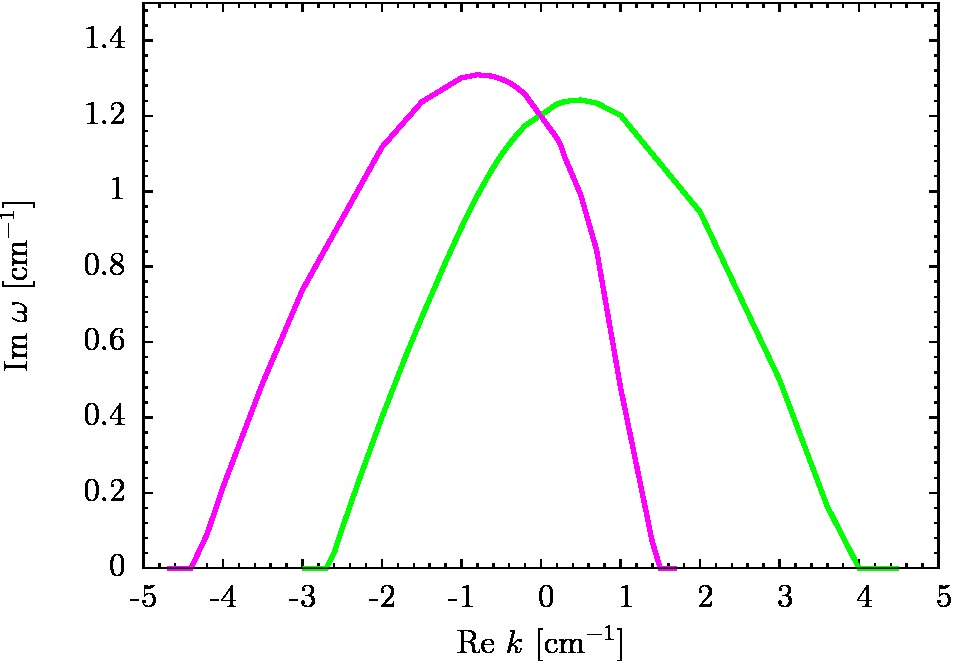}
\end{tabular} \caption{\label{growth} Growth rates of unstable modes as a function of $k$ for the radially inward direction (magenta) and the direction perpendicular to it (green). } 
\end{figure}

We now apply the linear analysis to this ELN. The dispersion relations (DR) between the absolute value of the wave number $k$ and the frequency $\omega$ of perturbations are displayed in the bottom panels of Fig. \ref{nu190} for different directions of $\mathbf k$ as indicated by arrows in the upper panels at $t_{pb}$ = 190 ms. All the patterns in these DRs given here were actually observed in our previous paper \cite{Azari:2019jvr} for artificially modified ELNs. This time the ELN is real, being obtained in our realistic self-consistent simulations. From the results in \cite{Azari:2019jvr} we can say that this  flavor state is unstable to the fast pairwise conversion and the growth rate will be largest for $\bold{k}$ directed radially inward, which is confirmed indeed in Fig. \ref{growth}. The instability is found for a much broader range of directions of $\mathbf {k}$ with similar growth rates, though, as understood from the same figure. The maximum growth rate is 1.3 $\mathrm{cm^{-1}}$ and the conversion occurs really fast compared with the hydrodynamical time scale. Note that the actual direction, in which perturbations grows, is not the direction of $\bold{k}$ but given by the group velocity. 

\begin{figure*}[t]
\begin{tabular}{ccc}
\includegraphics[width=4.5cm]{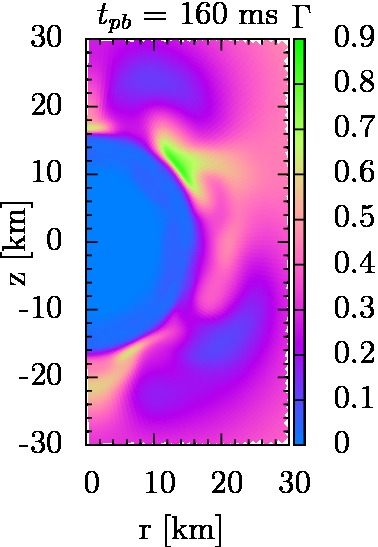}\includegraphics[width=4.5cm]{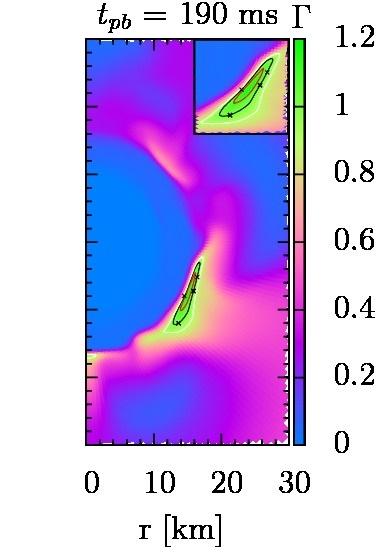}\includegraphics[width=4.5cm]{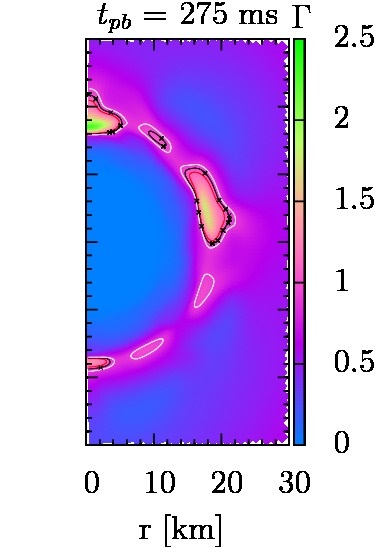}
\end{tabular} \caption{\label{Gamma-cross} The contour plots of the ratio of the number density of $\bar\nu_e$ to that of $\nu_e$ number densities, $\Gamma$, for the three different times. Black crosses for $t_{pb}$ = 190 and \mbox{275 ms} indicate the places where the crossings are observed. Three contour lines correspond to $\Gamma$ = 0.9 (white), 1.0 (black), and 1.1 (red). It is evident that the crossings occur when the value of $\Gamma$ is very close to 1. } 
\end{figure*}

Now we come to the core question: why the crossing occurs in this region of the core. In our previous paper we contended that the crossing is not easy to obtain, since the angular distributions of $\nu_e$ and $\bar{\nu}_e$ become more different as one goes deeper, where neutrinos are coupled with matter and their distributions are affected by matter motions, but the population of $\bar{\nu}_e$ will be suppressed by stronger Fermi-degeneracy of $\nu_e$ and vice versa at larger radii, where $\bar{\nu}_e$ is more abundant but the angular distributions become not much different between $\nu_e$ and $\bar{\nu}_e$. We find that in the regions, where the crossings occur, $\bar{\nu}_e$ is not suppressed and populated almost in comparable numbers to $\nu_e$ in fact.

In order to see this more quantitatively, we define $\Gamma$ as the ratio of the number density of $\bar\nu_e$, $n_{\bar\nu_e} = \int \frac {d^3p}{(2\pi)^3}f_{\bar\nu_e}$ to that of $\nu_e$, $n_{\nu_e} = \int \frac {d^3p}{(2\pi)^3}f_{\nu_e}$: 

\begin{equation}
\Gamma \equiv \frac{n_{\bar\nu_e}}{n_{\nu_e}}.
\end{equation}
Although we have investigated many snapshots at different post-bounce times, we show the color contour plots of $\Gamma$ for only three selected times ($t_{pb}$ = 160, 190 and 275 ms) in \mbox{Fig. \ref{Gamma-cross}}. We find that there are indeed some regions, where $\Gamma$ becomes close to unity near the PNS surface as early as $t_{pb}$ = 160 ms. It is clear the maximum value of $\Gamma$ for \mbox{$t_{pb}$ = 160 ms,} is 0.9 but it reaches to 1.2 and 2.5 at $t_{pb}$ = 190 and 275 ms, respectively. As we mentioned, we found the crossing for the first time at $t_{pb}$ $\sim$ 190 ms and we confirmed that the unstable regions continue to exist at later times. It seems indeed that the crossing occurs when $\Gamma$ reaches unity for the first time (see the middle panel of Fig. \ref{Gamma-cross}). Note that the contour lines in the second and third panels ($t_{pb}$ = 190 and 275 ms) in Fig. \ref{Gamma-cross} show $\Gamma$ = 0.9, 1.0 and, 1.1 respectively. It is seen that the crossing occurs (black cross signs ) when $\Gamma$ is very close to unity. This is understandable. Inside the neutrino sphere, the angular distributions of $\nu_e$ and $\bar{\nu}_e$ are both nearly isotropic (see Fig. \ref{ang190}) and, in order to get the crossing, their number densities should be almost the same. As mentioned above, \mbox{$\Gamma$} reaches unity for the first time at about $t_{pb}$ = 190 ms, when we find the crossings to happen. Although the maximum value of $\Gamma$ exceeds unity at later times, the ELN crossings still occur in the regions, where the value of \mbox{$\Gamma$ is $\approx$ 1} at these times. 

The widths of the unstable regions look very narrow but are still much wider than the conversion length scale, in fact. This is understood from Fig. \ref{Gamma-cross} as follows: the difference between the values of $\Gamma$ is $\Delta\Gamma$ = 0.1 for the adjacent contour lines that are seperated spatially by $\sim 0.5\times10^3$m; since the ELN is a fraction $\sim 10^{-3}$ of the individual number densities of $\nu_e$ and $\bar{\nu}_e$ in the region, where the ELN crossing occurs, the width of that region may be estimated as $\sim 5$m, which should be compared with the length of the fast conversion, $\sim$ 0.01 m (\mbox{see Fig. \ref{growth}}). 

\begin{figure*}[t]
\begin{tabular}{cc}
\includegraphics[width=8cm]{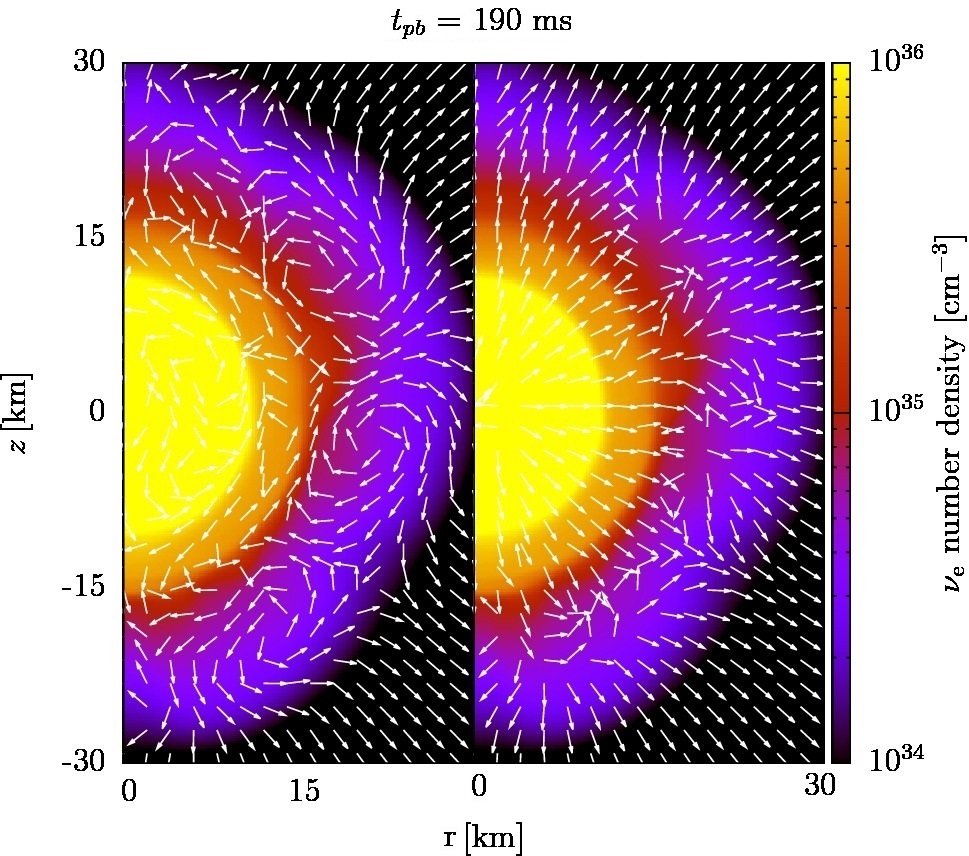}\includegraphics[width=8cm]{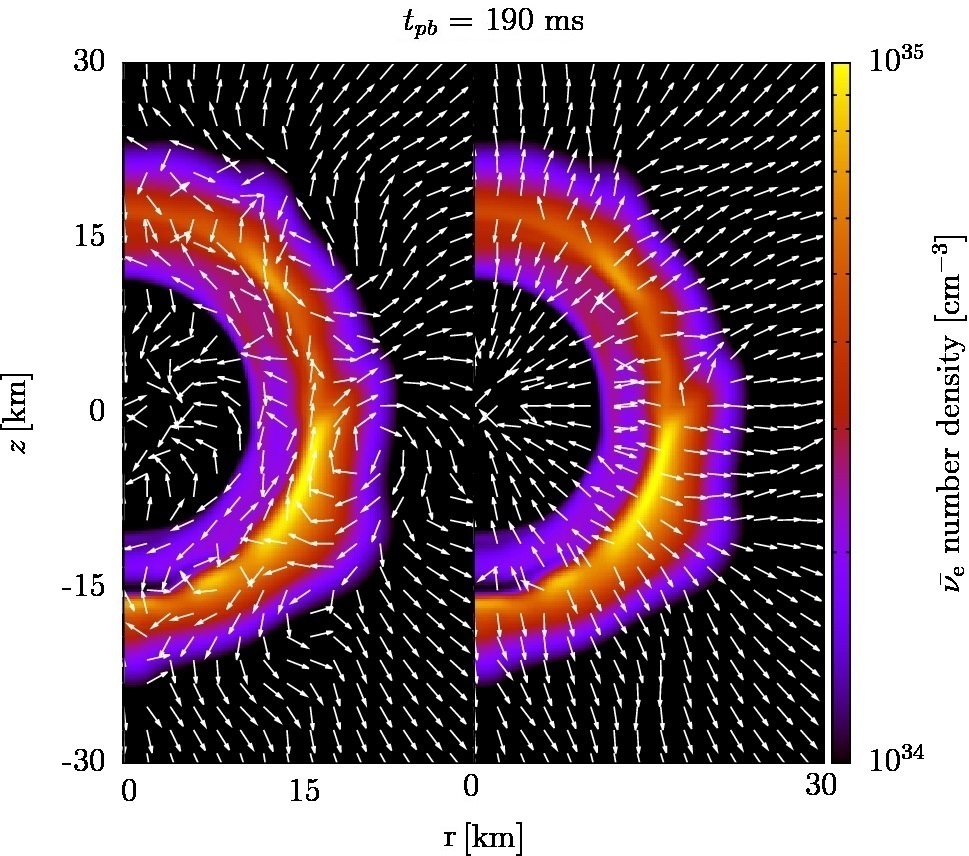}
\end{tabular} \caption{\label{flux} The number fluxes (arrows) and color contours of the number densities of $\nu_e$ (left two panels) and $\bar\nu_e$ (right two panels), respectively. The left and right halves of each panel are for the laboratory and fluid-rest frames, respectively.}
\end{figure*}

In Fig. \ref{flux} we present, instead of their ratios, the number densities of $\nu_e$ (left panel) and $\bar\nu_e$ (right panel) themselves at $t_{pb}$ = 190 ms as color contours. On top of them are the flux vectors of each species of neutrinos. In each panel, the left half is for the laboratory frame and the right half is for the fluid-rest frame. It is observed that $\nu_e$ becomes more abundant at smaller radii whereas $\bar{\nu}_e$ is almost absent at $r\lesssim$ 10 km. This is due to the strong Fermi-degeneracy of $\nu_e$ in this region. It is also obvious that $\bar{\nu}_e$ is most abundant off center at $r$ $\sim$ 15 km (see the yellow half-circles in the right panel of Fig. \ref{flux}). This  corresponds to the region, where the ratio $\Gamma$ is largest. The neutrino fluxes are trailing the convective matter motions irrespective of species in the laboratory frame as can be seen in the left half of each panel. In the fluid-rest frame, on the other hand, the fluxes are determined mainly by diffusions. As a result, the $\nu_e$ flux is mostly directed outwards, reflecting the fact that their number density is decreasing with radius rather monotonically. Some non-radial flows are recognized, though, in the regions, where the convective matter motions produce fluctuations of $Y_e$ (see Fig. \ref{ang190}) as well as of $n_{\nu_e}$ and $n_{\bar{\nu}_e}$; on the other hand, $\bar{\nu}_e$'s are diffusing inward at  $r\lesssim$ 15 km and outward at $r \gtrsim $15 km. This is just as expected, since their number densities are peaked at $r\sim$ 15 km as mentioned above. As a consequence of this fact, $\nu_e$ and $\bar{\nu}_e$ are flowing in the opposite directions at $r\lesssim$ 15km. This gives the crossings we found above. 

\begin{figure*}[t]
\begin{tabular}{ccc}
\includegraphics[width=4.5cm]{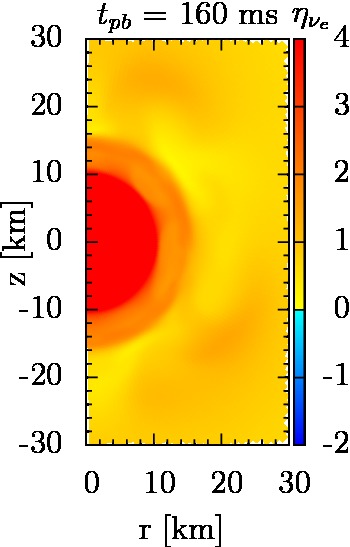}\includegraphics[width=4.5cm]{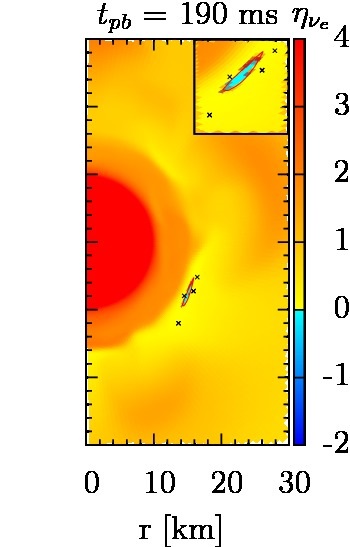}\includegraphics[width=4.5cm]{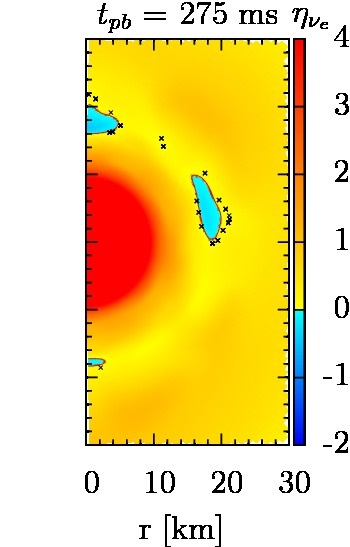} 
\end{tabular} \caption{\label{etamu} The color contour plots of $\eta_{\nu_e} = \mu_{\nu_e} / T$ at different times. Blue regions indicate the places, where $\eta_{\nu_{e}}$ is slightly smaller than 0 and the contour line indicates the points, at which $\eta_{\nu_{e}}$ = 0. The crosses show the points of ELN crossings. } 
\end{figure*}

We now consider why $\bar{\nu}_e$ has a peak in the number density at $r\sim$ 15 km. The crucial quantity is the chemical potential of $\nu_e$ defined as 

\begin{equation}
\mu_{\nu_e} = (\mu_{e} + \mu_{p}) - \mu_{n},
\end{equation}
where $\mu_e$, $\mu_p$ and $\mu_n$ are the chemical potentials of electrons, protons and neutrons, respectively. It is true that this has a rigorous thermodynamical meaning only when neutrinos are in chemical equilibrium, but the weak-equilibrium is almost established in this region in fact and hence $\mu_{\nu_e}$ serves us as a measure to gauge whether $\bar{\nu}_e$ is suppressed by Fermi-degeneracy or not. In Fig. \ref{etamu} we give the color maps of $\eta_{\nu_e} \equiv \mu_{\nu_e} / T$ at the same three post-bounce times. Crosses in this figure indicate again the places, at which we find the ELN crossings. One quickly recognizes that there appear a pocket of regions, where \mbox{$\eta_{\nu_e} \lesssim$ 0} (\mbox{light-bluish regions}), at $t_{pb} \sim 190$ ms and that the boundaries of these regions coincide with the regions we see the ELN crossings. Deeper inside, \mbox{$\eta_{\nu_e}$ $\gg$ 1} is satisfied and it is consistent with the previous observation that the presence of $\bar{\nu}_e$ is strongly suppressed.

\begin{figure}[h]
\begin{tabular}{c}
\includegraphics[width=8cm]{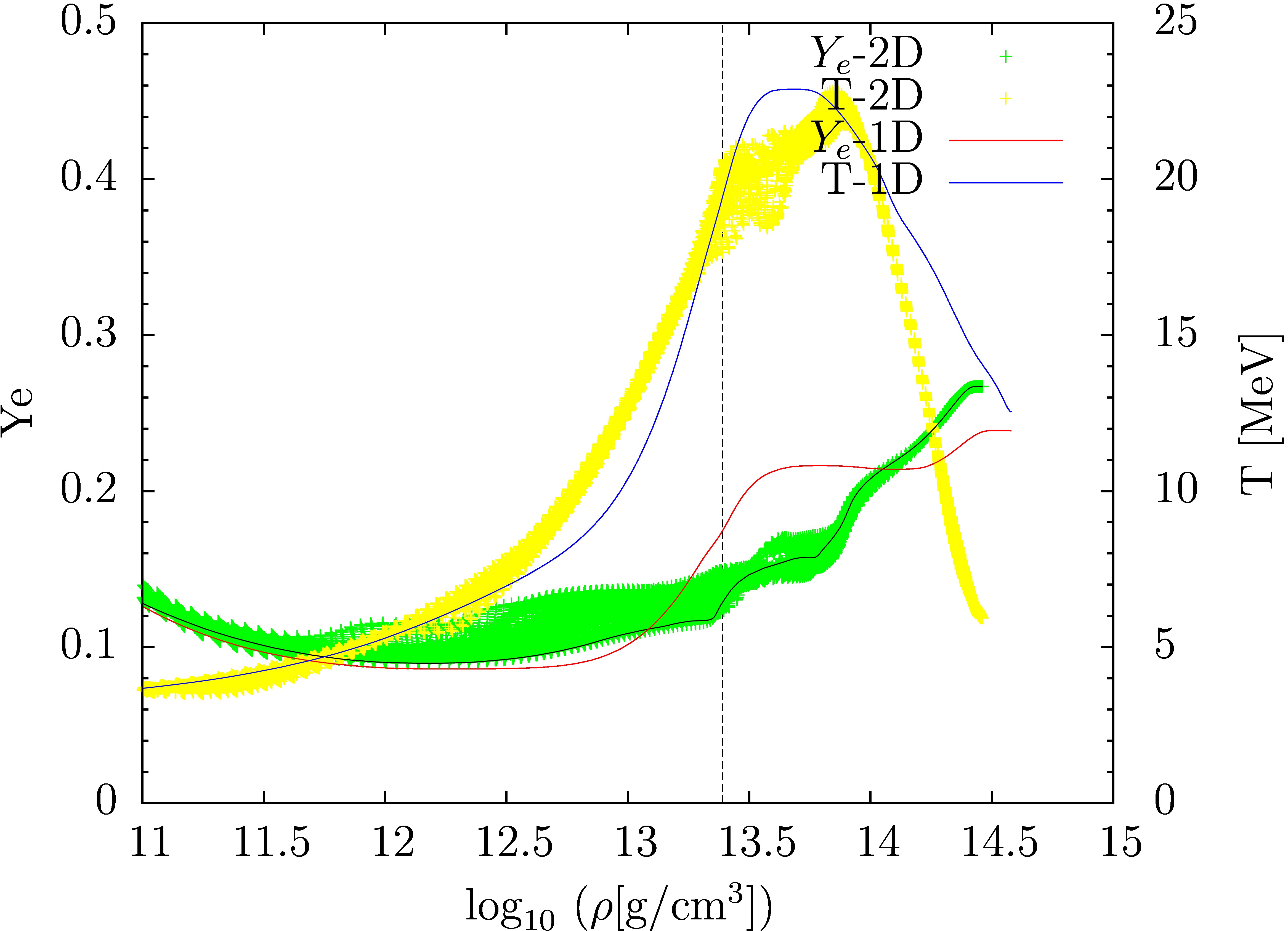}
\end{tabular} \caption{\label{1d-2d} Radial profiles of $T$ and $Y_e$ as a function of density for 1D (blue and brown lines for $T$ and $Y_e$, respectively.) and 2D (yellow and green points, respectively). The black line inside the 2D plots for $Y_e$ indicates the $Y_e$ profile along the radial ray in the direction of the ELN crossing. Black dashed line indicates the density, at which the ELN crossing is observed in 2D.} 
\end{figure}

Fig. \ref{1d-2d} shows the radial profiles of temperature $T$ and $Y_e$ as a function of density both for the present model and the 1D counterpart. Note that for the 2D case, we plot them for different radial directions. The results are a single line for 1D but a band (a superposition of many lines actually) for 2D with the width reflecting the fluctuations among different angles, for T and $Y_e$ individually. It is seen that the values of $Y_e$ in 1D is larger than the band for 2D around the region of our concern ($r$ $\sim$ 16.5 km and $\rho$ $\sim$ 2.4 $\times 10^{13}\mathrm{g/cm^3}$); in particular, the values of $Y_e$ in the direction, in which the ELN crossing occurs in 2D, are running along the bottom of the band except for $\log(\rho \mathrm{[g/cm^3]})$$\gtrsim$13.5.
Note that it is known that the PNS convection facilitates the deleptonization of PNS \cite{Dessart:2005ck,Buras:2005tb}. It should be mentioned that the above 1D model was computed not with the Furusawa-Shen EOS (FSEOS) as for the 2D model but with the Furusawa-Togashi EOS (FTEOS)\cite{Furusawa:2017auz} based on the variational method with realistic nuclear potentials, since that is the only model for the same progenitor available to us at the moment. As we will show later at the end of this section, this will not be a concern.

\begin{figure}[h]
\begin{tabular}{c}
\includegraphics[width=8cm]{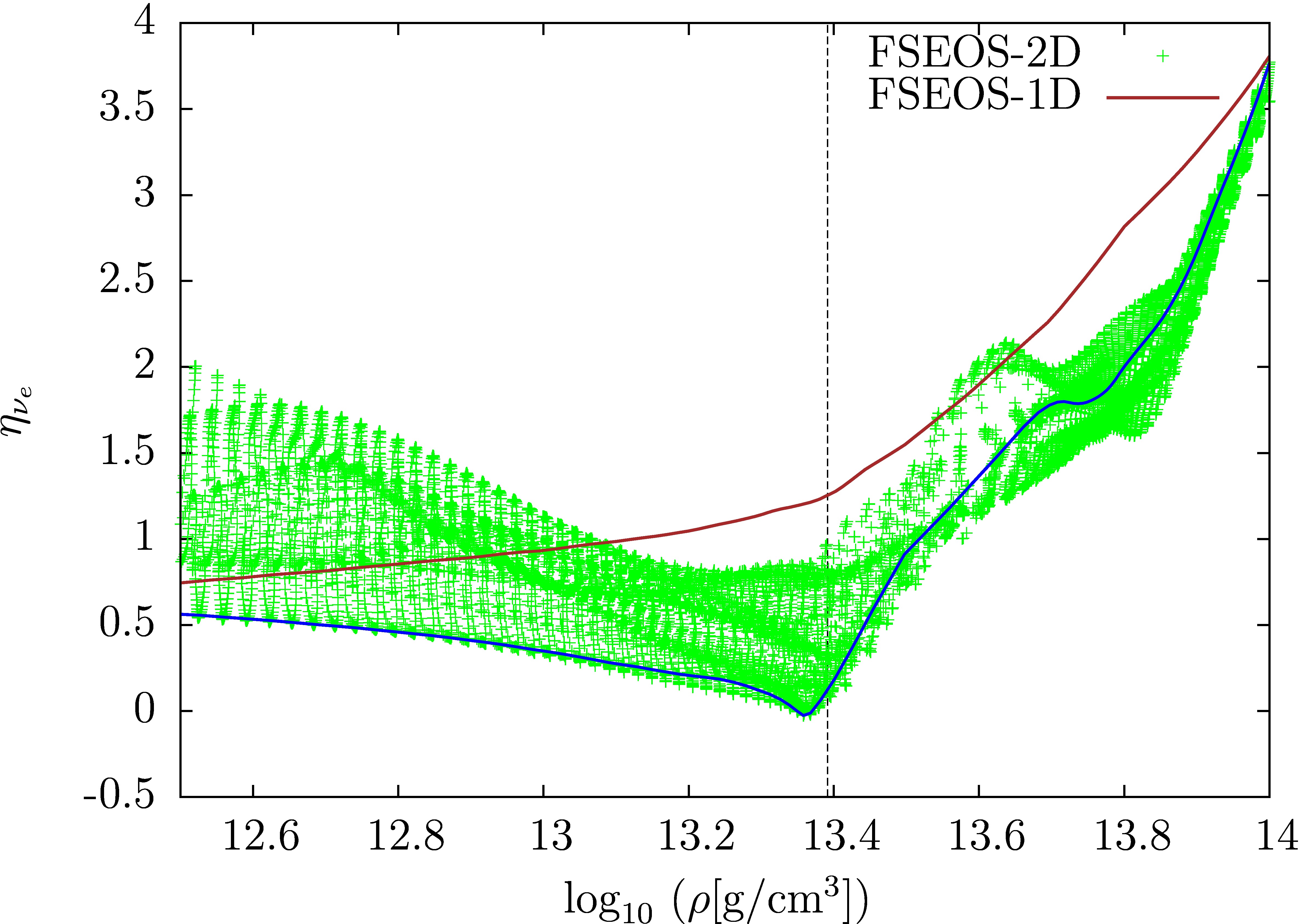}
\end{tabular} \caption{\label{eoss1} Radial profile of the degeneracy parameter of electron-type neutrino $\eta_{\nu_e}$ as a function of density for 1D and 2D. Blue solid line corresponds to the profile of $\eta_{\nu_e}$ along the radial ray in the direction of the ELN crossing. Black dashed line indicates the density of our focus.} 
\end{figure}

The fact that $\nu_e$ and $\bar{\nu}_e$ are flowing in the opposite directions at small radii ($r \lesssim$ 10 km) is well known in the proto-neutron star (PNS) cooling. Since this happens in 1D already, the multi-dimensional effect should be important for the ELN crossing, which is not observed in 1D. 
In Fig. \ref{eoss1}, we present the degeneracy parameter of $\nu_e$ again as a function of density for both 1D and 2D. We find that the degeneracy of $\nu_e$ is not resolved ($\eta_{\nu_e}\sim$ 1) in 1D although it is lowered substantially in the region of our concern ($r$ $\sim$ 16.5 km and $\rho$ $\sim$ 2.4 $\times 10^{13}\mathrm{g/cm^3}$) unlike in 2D; as a result, the ratio $\Gamma$ never comes close to unity in 1D, implying no chance for ELN crossing.

\begin{figure}[t]
\begin{tabular}{c}
\includegraphics[width=8.5cm]{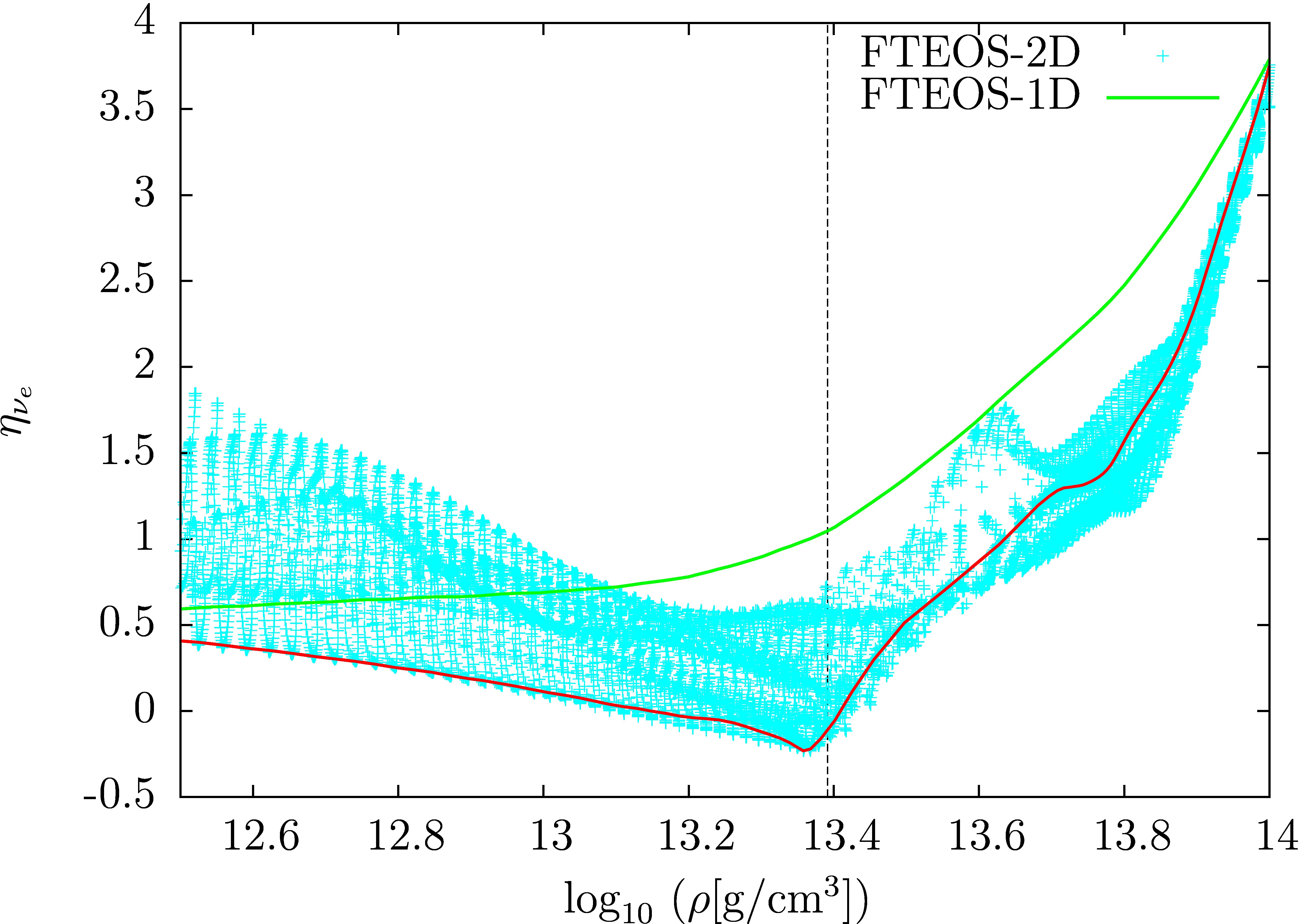} 
\end{tabular} \caption{\label{eoss} Same as Fig. \ref{eoss1} but for FTEOS.} 
\end{figure}

To give some numbers of relevance for one of the representative points (\mbox{$r$ = 16.5 km} and \mbox{$\theta$ = 2.1 rad}), where detailed analyses are done so far, the values of denisty, temperature and electron fraction are  \mbox{$\rho$ = 2.4 $\times 10^{13}\mathrm{g/cm^3}$}, $T$ = 20.4 MeV and $Y_e$ = 0.13, respectively, and the chemical potentials of electron, proton and neutron including the rest mass are $\mu_e$ = 56.9 MeV, $\mu_p$ = 848 MeV and $\mu_n$ = 903 MeV, respectively. The electron is hence strongly degenerate here as expected. It is also found, however, that this large chemical potential of electron is almost canceled by the chemical potential difference between proton and neutron. 

The mass fraction of free neutron and proton are $X_n$ = 0.781 and $X_p$ = 0.069, respectively. If they were non-interacting ideal Boltzmann gases, they would have smaller chemical potential difference by $\sim$ 7 MeV than those given above. Nuclear interactions are hence playing roles here. What is more important is  the existence of light elements in the region. In fact, the mass fraction of light nuclei (Z $<$ 6) is $X_l$ = 0.138 whereas that of heavy elements (Z$\geq$6) is $X_h$ = 1.16 $\times 10^{-2}$ at the point of our current concern. Since free protons are largely absorbed in the light elements, they become less abundant and the chemical potential difference between neutron and proton is enhanced. 

The existence of the light elements in the region of our concern is observed also in the 1D model, however. In fact, the mass fractions of relevance are $X_n$ =0.726, $X_p$ = 0.098, $X_l$ = 0.173 and $X_h$ = $2.1 \times 10^{-4}$. It is noted again that this 1D model was computed with the Furusawa-Togashi EOS (FTEOS)\cite{Furusawa:2017auz} but the mass fractions are recalculated in post-process with the FSEOS for the density, temperature and $Y_e$ obtained in the simulation. As will be shown shortly in the next paragraph, this inconsistency is not a serious problem.
It turns out that these numbers do not give $\eta_{\nu_e} \sim 0$ (see Fig. \ref{eoss1}). The critical difference comes from $Y_e$. As mentioned above, the 2D model realizes smaller values of $Y_e$ in the direction of the ELN crossing owing to convective matter motions. In the 1D model, the higher values of $Y_e$ owing to the absence of such non-radial motions give higher values of $X_p$ and lower values of $X_n$, both of which tend to reduce the chemical potential difference.
The result is the larger value of the chemical potential of $\nu_e$ and the smaller value of $\Gamma$, which suppress the ELN crossing, as mentioned earlier.
It is now apparent that the combination of the smaller values of $Y_e$ induced by convective matter motions and the emergence of the light elements is the ultimate cause of the ELN crossing we found in this paper.

Finally, in Fig. \ref{eoss} we show $\eta_{\nu_e}$ obtained with the Furusawa-Togashi EOS (FTEOS) \cite{Furusawa:2017auz}, which is based on the variational method with realistic nuclear potentials and satisfies various constraints for the hadronic EOS well and is hence favored nowadays. It is clear that the values of $\eta_{\nu_e}$ are not much different from those derived with the FSEOS. Indeed, we can see $\eta_{\nu_e}$ $\sim$ 0 at almost the same point. This suggests that the EOS-dependence is rather weak although more work is certainly needed.

\section {Summary and Discussions}

We have extended our previous pilot study and conducted a full survey of the post-bounce core of \mbox{11.2 $M_{\odot}$} progenitor, searching for the \mbox{electron lepton number (ELN)} crossing. We have reported positive detections of the crossings at \mbox{$r \sim$ 16 - 21 km} from the post-bounce time of \mbox{$t_{pb}$ = 190 ms} onward. These regions are located inside the neutrino sphere and may have an impact on the CCSNe explosion. Conducting linear analysis for the representative points and times, we have confirmed that the crossing really induces the fast flavor conversion at least in the linear level and that the conversion occurs very quickly. As a result, although the regions, where we have found the crossings, are confined to narrow shells, they are still wide enough for the conversion to occur. 

\begin{figure*}[t]
\begin{tabular}{c}
\includegraphics[width=10cm]{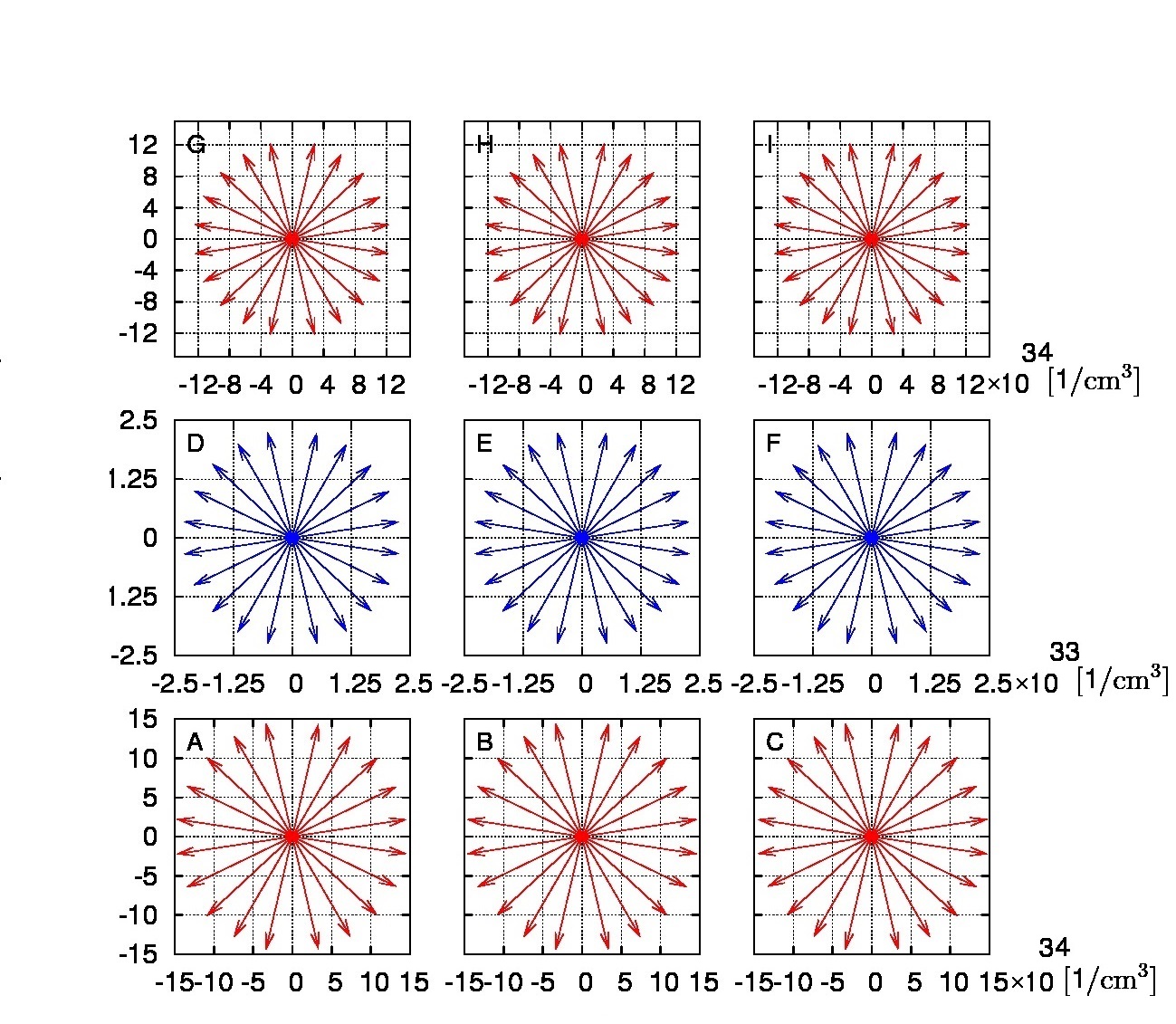} 
\end{tabular} \caption{\label{ang160} Same as Fig. \ref{ang190} but for $t_{pb}$ = 160 ms. Note that the scales are different for $\nu_e$ (red) and $\bar{\nu}_e$ (blue).} 
\end{figure*}

\begin{figure*}[t]
\begin{tabular}{cc}
\includegraphics[width=6cm]{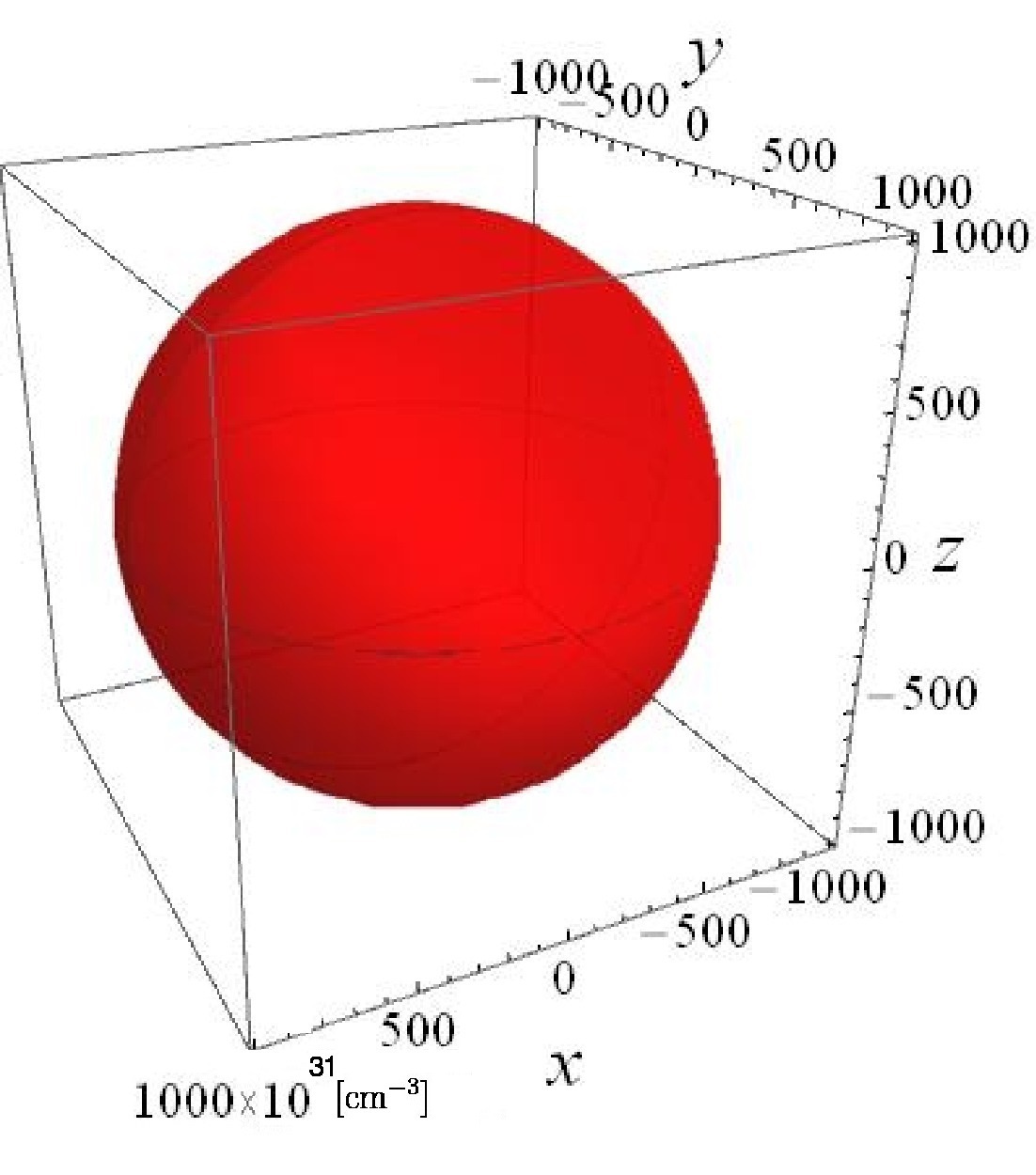}\includegraphics[width=6cm]{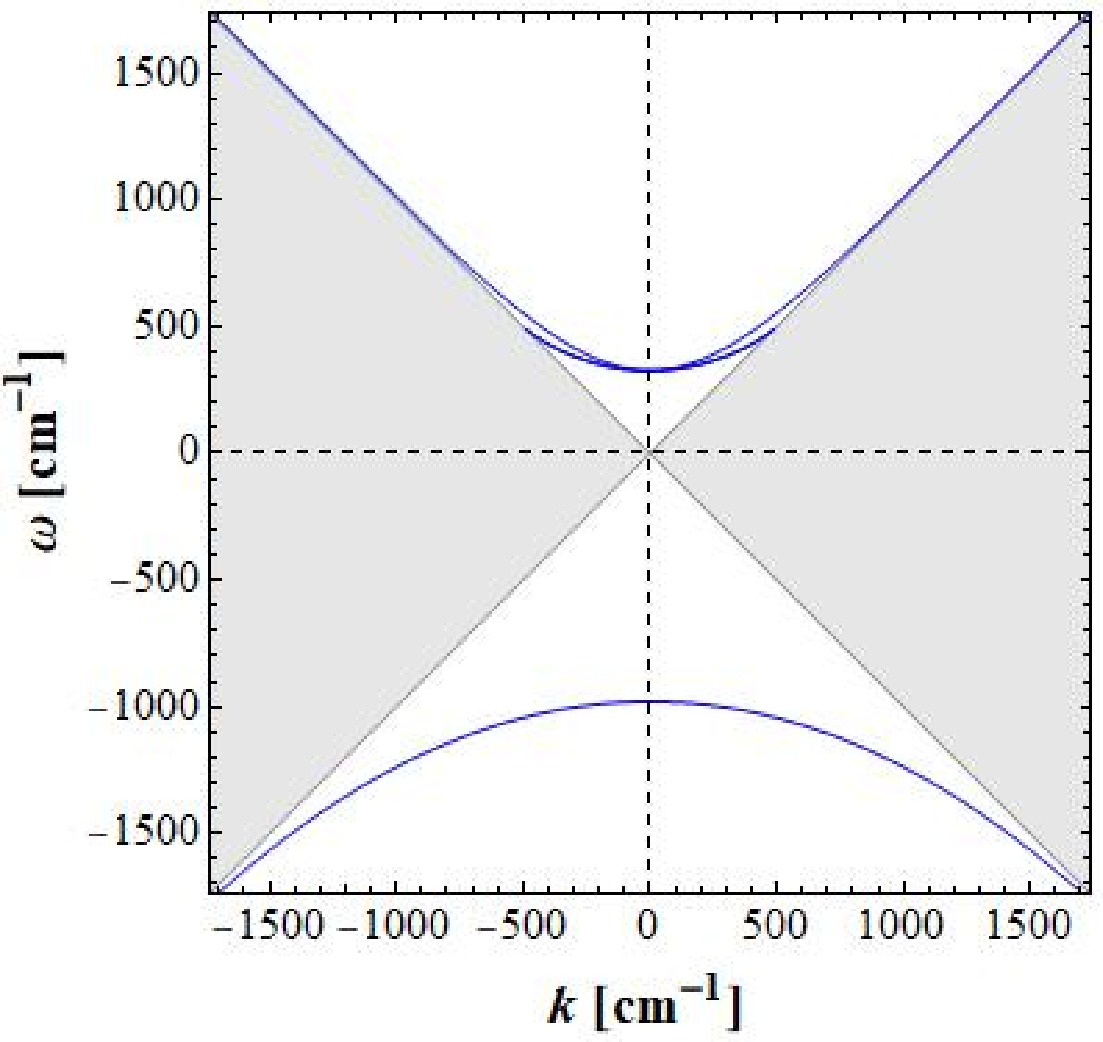} 
\end{tabular} \caption{\label{nu160} Same as Fig. \ref{nu190} but for $t_{pb}$ = 160 ms. The red color indicates that the $\nu_e$ is dominant over $\bar\nu_e$.}
\end{figure*}
We have then studied rather in detail why the crossing is obtained at these regions alone. We have observed that the population of $\bar{\nu}_e$ is comparable to or even larger than that of $\nu_e$ in those regions and that the ratio $\Gamma$ of the former to the latter tends to increase in time. Since the angular distributions of $\nu_e$ and $\bar{\nu}_e$ are both nearly isotropic, the ELN crossing is possible only where $\Gamma$ is very close to 1, which is really the case. 

We have found that the chemical potential of $\nu_e$ becomes almost vanishing although the electrons are still strongly degenerate and demonstrated that the large chemical potential of electron is almost canceled by the chemical potential difference of neutron and proton. 
This became possible by the interplay of nuclear physics and hydrodynamics, i.e., the appearance of light elements, which take in most of the free protons, and lower values $Y_e$ produced by the convective motions of matter. This phenomenon seems to depend on EOS rather weakly and is may occur commonly. 

It should be mentioned finally that the fact that we have found the ELN crossings and hence the fast flavor conversions in some regions inside the neutrino sphere is one thing but whether they have really some implications for supernova explosions is quite another. For one thing, the solid angle that these sporadically regions subtend is not very large although it tends to widen in time. We had better look into other simulation results for different progenitors and/or with different EOS's. Axisymmetry we imposed in our simulations should be removed. All three flavors should be taken into account. The nonlinear evolution of the fast flavor conversion should be investigated \cite{Richers:2019grc,Abbar:2018beu}. And eventually we need to perform CCSN simulations with these fast flavor conversions begin somehow implemented \cite{Chakraborty:2019wxe}.  

\begin{acknowledgments}

M.D.A was supported by the Ministry of \mbox{Education}, Culture, Sports, Science and Technology of Japan (MEXT) and Waseda University for his \mbox{post-graduate studies.} T.M is supported by JSPS Grant-in-Aid for JSPS Fellows (No.19J21244) from the Ministry of Education, Culture, Sports, Science and Technology (MEXT), Japan. H.N was supported by Princeton University through DOE SciDAC4 Grant DE-SC0018297(subaward 00009650). A.H is partially supported by MEXT Grant-in-Aid for Research Activity Start-up (No.19K23435). K.S is partially supported by Grant-in-Aid for Scientific Research (26104006, 15K05093, 19K03837). This work is supported by Waseda University grant for special research projects (2019R-041). 
This work is also supported by Grant-in-Aid for Scientific Research (19H05811, 26104006, 15K05093) and Grant-in-Aid for Scientific Research on Innovative areas "Gravitational wave physics and astronomy:Genesis" (17H06357, 17H06365) from the Ministry of Education, Culture, Sports, Science and Technology (MEXT), Japan.
 For providing high performance computing resources, Computing Research Center, KEK, JLDG on SINET4 of NII, Research Center for Nuclear Physics, Osaka University, Yukawa Institute for Theoretical Physics, Kyoto University, Nagoya University, and Information Technology Center, University of Tokyo XC30 and the general common use computer system at the Center for Computational Astrophysics, CfCA, the National Astronomical Observatory of Japan are acknowledged.
This work was supported by HPCI Strategic Program of Japanese MEXT and K computer at the RIKEN (Project ID: hpci 160071, 160211, 170230, 170031, 170304, hp180179, hp180111), "Priority Issue on Post-K computer" (Elucidation of the Fundamental Laws and Evolution of the Universe) and Joint Institute for Computational Fundamental Sciences (JICFus).
 The numerical computations were performed on the K computer, at AICS, FX10 at the Information Technology Center of the University of Tokyo.

\end {acknowledgments}

\appendix*
\section{}

We present the results of our analysis for \mbox{$t_{pb}$ = 160 ms} as a representative of earlier post-bounce times, at which we find no ELN crossing for the $\nu_e$ and $\bar\nu_e$ angular distributions. As demonstrated in our previous paper \cite{Azari:2019jvr} in the early post-bounce phase $\nu_e$ are substantially more abundant than $\bar\nu_e$. We pick up the same point (\mbox{$r$ = 16.5 km} and \mbox{$\theta$ = 2.1 rad}) where the ELN crossing is detected at $t_{pb}$ = 190 ms and we gave detailed analyses in the main body of the paper. \mbox{In Fig. \ref{ang160}} we show the angular distributions of $\nu_e$ and $\bar\nu_e$ (cf. Fig. \ref{ang190}). It is clear that the scales are different between $\nu_e$ and $\bar\nu_e$  and $\nu_e$ is highly dominant over $\bar\nu_e$ in all directions.

For comparison, we give some relevant numbers: the density, temperature and electron fraction are \mbox{$\rho$ = 2.36 $\times 10^{13} \mathrm{g/cm^3}$}, $T$ = \mbox{18.4 MeV} and $Y_e$ = 0.1, respectively; the mass fractions are $X_n$ = 0.827, $X_p$ = 0.051, $X_h = 4.58\times10^{-3}$, and $X_l$ = 0.116. The chemical potentials of electron, proton and neutron are $\mu_e$ = 66.5, $\mu_p$ = 861 and \mbox{$\mu_n$ = 907 MeV} including rest masses, respectively. Since we obtain \mbox{$\eta_{\nu_e} \gtrsim $ 1}, it is natural that $\bar{\nu}_e$ is suppressed. 
\mbox{Fig. \ref{nu160}} shows the angular distributions difference between $\nu_e$ and $\bar\nu_e$ (left panel) for $\mathbf k$ oriented in the radial direction. Note that the red color means that the $\nu_e$ is dominant over $\bar\nu_e$ (cf. Fig. \ref{nu190}) and the corresponding  dispersion relation DR (right panel). The DR has a pattern typically observed when there is no ELN crossing \cite{Azari:2019jvr} where branches that have a gap in the frequency of perturbation $\omega$.  

\bibliography{Papers_vf}

\end{document}